\documentclass[twocolumn,aps]{revtex4}

\usepackage{graphicx}
\usepackage{dcolumn}
\usepackage{bm}
\usepackage{psfig}

\begin{document}
\title{Magnetic interaction between GaMnAs  layers via spin-polarized quasi-two-dimensional
hole gas: Monte Carlo simulation\\
Session BB-6}
\author{M. A. Boselli, L. Loureiro da Silva, I. C. da Cunha Lima}
\address{Instituto de F\'\i sica, Universidade do Estado do Rio de Janeiro\\
Rua S\~{a}o Francisco Xavier 524, 20.500-013 Rio de Janeiro, R.J., Brazil}
\author{A. Ghazali}
\address{Groupe de Physique des Solides, Universit\'{e}s Paris 7 et Paris 6\\
Tour 23, 2 Place Jussieu, F-75 125 Paris Cedex 05, France}

\date{\today}

\begin{abstract}
The magnetic order resulting from an indirect exchange between
magnetic moments in the metallic phase of a Ga$_{1-x}$Mn$_{x}$As
bilayered structure is studied \textit{via} Monte Carlo
simulation. The coupling mechanism involves a perturbative
calculation in second order of the interaction between the
magnetic moments and carriers (holes). We take into account a
possible polarization of the hole gas establishing, thus,
self-consistency between the magnetic order and  the electronic
structure. It leads to a ferromagnetic order even in the case of
thin layers. This fact is analyzed in terms of the inter- and
intra-layer interactions.
\end{abstract}

\pacs{75.70.-b, 75.10.-i, 75.70.Cn}
\maketitle

The possibility of having a diluted magnetic semiconductor (DMS)
based on GaAs opens a wide range of potential applications as, for
instance, in integrated magneto-optoelectronic devices \cite{nat}.
In Ga$_{1-x}$Mn$_{x}$As alloys substitutional Mn acts as an
acceptor (it binds one hole), and at the same time it carries a
localized magnetic moment, due to its five electrons in the
\textit{3d} shell. For $x=0.053$, the alloy is a metallic
ferromagnet \cite{matsukura}, the Curie-Weiss temperature is 110K,
and the free hole concentration is near $10^{20}cm^{-3}$. The
ferromagnetic order in the metallic phase is understood, at
present, as resulting from the indirect exchange between the
Mn$^{2+}$ ions due to the local spin polarization in the hole gas.
This explanation implies the spin coherence length to be larger
than the average distance of the localized moments.

In this work we extended a confinement-adapted RKKY \cite{bos01}
mechanism  to study the magnetic order resulting of the indirect
exchange between magnetic moments in a GaAs/Ga$_{1-x}$Mn$_{x}$As
nanostructure with two DMS layers. A Monte Carlo simulation is
performed to determine the resulting magnetic phases. The
interaction potential between a Fermi gas and a set of localized
magnetic moments is well described by the Hund-type exchange
potential:

\begin{equation}
H_{\text{ex}}=-I\sum_i\vec S_i\cdot \vec s(\vec r)\delta(\vec r- \vec R_i),
\label{hund}
\end{equation}
where $\vec S_i$ is the localized spin of the Mn ion at position
$\vec R_i$, which will be treated as a classical variable, and
$\vec s(\vec r) $ is the spin operator of the fermion at $\vec r$;
$I$ is the $sp-d$ interaction \cite{szcz}. In terms of the hole
field operators, the Hund Hamiltonian is
\[
H_{\text{ex}}=-\frac{I}{2}\sum_{i}\{S_{i}^{z}[\psi_{\uparrow}^{\dagger}(\vec{r}_i)
\psi_{\uparrow}(\vec{r}_i)-\psi_{\downarrow}^{\dagger}(\vec{r}_i)
\psi_{\downarrow}(\vec{r}_i)]
\]
\begin{equation}
+
S_{i}^{+}\psi_{\downarrow}^{\dagger}(\vec{r}_i)\psi_{\uparrow}(\vec{r}_i)+
S_{i}^{-}\psi_{\uparrow}^{\dagger}(\vec{r}_i)\psi_{\downarrow}(\vec{r}_i)\}
\label{hexfield}
\end{equation}
Instead of free fermions in a 3-D space, the electrons and holes
in a semiconductor heterostructure are confined in the growth
direction, assumed to be the z-axis, due to the mismatch of the
conduction and valence band edges. The total Hamiltonian,
$H=H_0+H_{\text{ex}}$ includes the kinetic part, the confinement
potential and the Hartree as well as exchange and correlation
terms in $H_0$ \cite{luc}.

Neglecting the scattering by impurities, holes are free particles
in the plane perpendicular to that growth direction, i.e., in the
plane parallel to the semiconductor interfaces. Their field
operator can be written as

\begin{equation}
\hat{\psi}_{\sigma }(\vec{r})=\frac{1}{\sqrt{A}}\sum_{n,\vec{k}}e^{i\vec{k}.%
\vec{\rho}}\phi _{n,\sigma}(z)\eta_{\sigma} c_{n,\vec{k},\sigma },
\label{field}
\end{equation}
where $A$ is the normalization area, $\vec{k}$ is a wave vector in the plane (%
$x,y$), $\eta _{\sigma}$ is the spin tensor for the polarization
$\sigma$, $\phi _{n,\sigma}(z)$ is the envelope function
which describes the motion of the fermion in the $z$-direction, and $c_{n,%
\vec{k},\sigma }$ is the fermion annihilation operator for the state ($n,%
\vec{k},\sigma $). $\vec{\rho}$ represents a vector in the 2-D
coordinates plane ($x,y$).

The confined RKKY indirect exchange is a second order perturbative
treatment \cite{bos01}.  This approach can be extended to start
the perturbative calculation from spin-polarized states, going, in
one sense, beyond the second order of perturbation. In that case,
we do not obtain a simple scalar product of localized dipole
moments in a Heisenberg-like Hamiltonian, since the polarization
breaks the rotational symmetry establishing a preferential
direction, that of the average magnetization. After a lengthy
calculation, however, it can be shown that the interaction term
can be written as an effective Hamiltonian:
\[
H_{eff}=-\sum_{i,j}
(C_{ij}^{\uparrow\uparrow}+C_{ij}^{\downarrow\downarrow})S_i^zS_j^z+
\]
\begin{equation}
(C_{ij}^{\uparrow\downarrow}+C_{ij}^{\downarrow\uparrow})(S_i^xS_j^x+S_i^yS_j^y)
\label{newrkky}
\end{equation}
where $x,y$ and $z$ (the later assumed to be the direction of
magnetization) are spin coordinates, and $i$ and $j$ are indices
of the magnetic moment. The $sp-d$ interaction, is generally
written as $I/v_{0}=N_{0}\beta $, where $v_{0}$ is the effective
volume of the  $Mn$ ion ($a_{0}^{3}/4x$ in the present case),
$N_{0}$ is the density of atoms of $Ga$, and $\beta $ is the
$sp-d$ exchange. For $GaMnAs$, $N_{0}\beta =-1.2$eV \cite{okaba}.
The exchange coefficients $C_{ij}^{\mu\nu}$ are expressed in terms
of the real space Fourier transform of the Lindhard function:
\[
C_{ij}^{\mu\nu}=-\sum_{n\in\mu}\sum_{n'\in\nu}\sum_{\textbf{q}}
(\frac{I}{2A})^2
\phi^*_{n'}(z_i)\phi_{n}(z_i) \times
\]
\begin{equation}
\times
\phi^*_{n}(z_j)\phi_{n'}(z_j)
\chi^{n,n^{\prime}}(\textbf{R}_{ij}) \label{defcij}
\end{equation}
with

\begin{equation}
\chi^{n,n^{\prime}}(\textbf{R}_{ij})=\sum_{\vec k}\exp [-i\vec
q\cdot \textbf{R}_{ij}]\chi ^{n,n^{\prime }}(\vec q)
\end{equation}
The Lindhard function is defined by \cite{keld}:
\begin{equation}
\chi ^{n,n^{\prime }}(\vec q)=\sum_{\vec k} \frac{\theta
(E_F-\epsilon _{n,\vec k})-\theta (E_F-\epsilon _{n^{\prime },
\vec k+\vec q})}{\epsilon _{n^{\prime },\vec k+\vec q}-\epsilon
_{n,\vec k}}.  \label{modlin}
\end{equation}

\begin{figure}[tbp]
\psfig{file=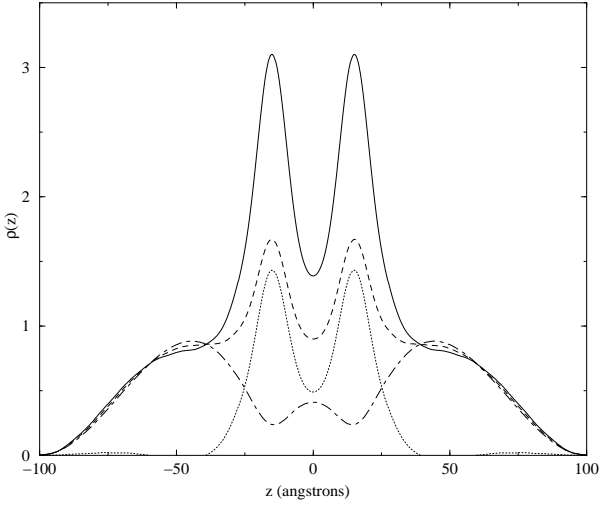,width=8cm,angle=270}
\caption{ Spin-polarized charge density distributions for two
Ga$_{0.95}$Mn$_{0.05}$As layers of 10\AA\  , separated by a 20\AA\
width $GaAs$ layer, and hole concentration $p=1\times
10^{20}$cm$^{-3}$. Solid line for total charge density, dotted
line for spin $\uparrow$ (parallel), dashed line for spin
$\downarrow$ (anti-parallel). The polarization density in
dotted-dashed line.}
\end{figure}

We have focused our attention on bilayered systems, consisting of
two GaMnAs layers inside an otherwise  nonmagnetic GaAs matrix.
Fig.(1) shows the density of holes, charge density for each spin
orientation, and the polarization density for the case where two
10\AA\  GaMnAs layers separated by 20\AA\  have their magnetic
moments fully aligned. It results from a self-consistent
calculation \cite{luc} of the eigenstates of $H_0$ with $x=5\%$
and the hole concentration is $10\times 10^{20}$cm$^{-3}$. The
interaction, given by Eq.(\ref{newrkky}), is assumed to be
effective within a cutoff radius, which we have taken as
$R_{c}=2.5a$, where $a$ is the fcc lattice parameter of GaAs. The
calculation is performed in a finite box, whose axes are parallel
to the [100] directions. Its dimensions are $L_{x}=L_{y}$, and
$L_{z}=Na/2$, and $N$ is the number of DMS monolayers (ML).
Periodic boundary conditions are imposed in the $(x,y)$ plane. The
lateral dimensions are adjusted in such a way that the total
number $N_{s}$ of spins ranges from 3500 to 4100, for all samples
with different $L_{z}$. The initial spin orientations are randomly
assigned. At a given temperature, the energy of the system due to
the interaction is calculated, and the equilibrium state for a
given temperature is sought by changing the individual spin
orientation according to the Metropolis algorithm. A slow cooling
stepwise process is accomplished making sure that the thermal
equilibrium is reached at every temperature. The resulting spin
configuration is taken as the starting configuration for the next
step at a lower temperature. For every temperature, the average
magnetization $<M>$ is calculated. Whenever the magnetization
reaches thresholds of multiples of tenths of $5\hbar/2$, the
electronic structure is recalculated assuming a homogeneous
magnetization in the DMS layers, and the $C_{ij}^{\mu\nu}$
coefficients are also recalculated according to Eq.
(\ref{defcij}). In a sense, this approach establishes a
self-consistency in the calculation of the magnetic order and in
the electronic structure.

\begin{figure}[tbp]
\psfig{file=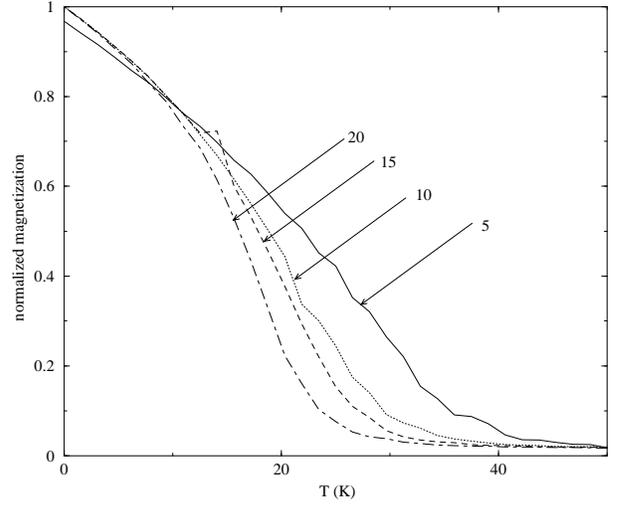,width=8cm,angle=270}
\caption{Magnetization as a function of temperature for two
Ga$_{0.95}$Mn$_{0.05}$As layers of 10\AA\  width and hole
concentration $p=1\times 10^{20}$cm$^{-3}$, separated by a $GaAs$
layer whose width in \AA\  is indicate by arrows.}
\end{figure}

We have tested the influence of the separation of the two DMS
layers. It is known that a ferromagnetic order can not occur in a
single DMS layer below a certain layer width \cite{bos01}.
However, the magnetic order in a DMS bilayer responds to the
interaction occurring both intra- and inter- layer. Fig.(2) shows
the magnetization as a function of temperature, obtained for DMS
thickness of 10 \AA\  in each layer, with a hole concentration of
$ p= 1 \times 10^{20}$ cm$^{-3}$, for different layer separation.
We have chosen an interaction cutoff radius of 5 monolayers, that
is greater them the mean Mn distance, and large enough to take
account for anti-ferromagnetic interactions. Increasing the
inter-layer separation, the Curie temperature decreases as a
direct consequence of the fact that the inter-layer interaction
decreases with the separation. Fig.(3) shows results for a similar
calculation, but $ p= 2 \times 10^{20}$ cm$^{-3}$. We observe a
rising on the Curie temperature. Also, increasing the carrier
concentration implies in increasing the Fermi wave vector what
introduces the possibility of anti-ferromagnetic interactions
within the chosen cutoff radius. This produces a partial
magnetization in the curves, associated with the occurrence of
canted phases \cite{bos01}.

\begin{figure}[tbp]
\psfig{file=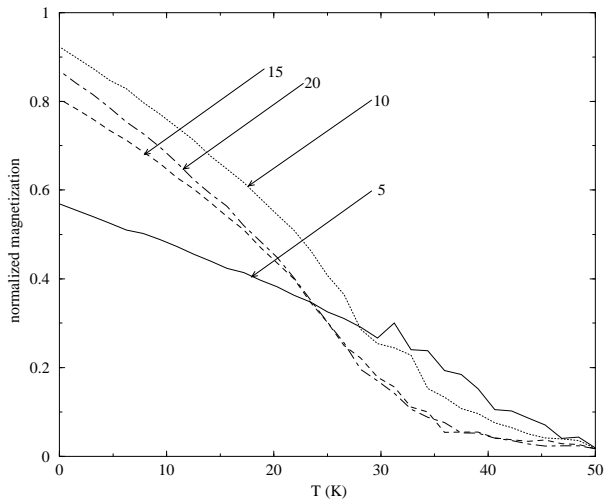,width=8cm,angle=270}
\caption{Same as above for $p=2\times 10^{20}$cm$^{-3}$}
\end{figure}

Our results show that a ferromagnetic order with significant Curie
temperature occurs in bilayered structure, even in the case of
thin DMS layers. For the sake of completeness we calculated also
some samples with two DMS layers of 20 \AA\  each. On these
systems we observed the same behavior described above, but with
higher Curie temperatures as expected for thicker layers.

\acknowledgements This work was supported by  CNPq and FAPERJ in
Brazil, and by the CAPES (Brazil) - COFECUB (France)Program.

\end{document}